\def\figsize{9.0cm}

\def\rn{\noindent\parshape 2 0truecm 8.8truecm 0.3truecm 8.5truecm}
\def\nn#1 #2{#1, #2.}				
\def\nnn#1 #2 #3{#1, #2. #3.}			
\def\nnnn#1 #2 #3 #4{#1, #2. #3. #4.}		
\def\nnnnn#1 #2 #3 #4 #5{#1, #2. #3. #4. #5.}	
\def\dualand{, \&\hbox{ }}				
\def\multiand{, \&\hbox{ }}				

\def\rg#1;#2;#3;#4;#5;#6 {\par\rn#1 #2, {\it #3}, {\bf #4}, #5 (``#6'') \par}
\def\rf#1;#2;#3;#4;#5 {\par\rn#1 #2, {\it #3}, {\bf #4}, #5\par}
\def\rfbook#1;#2;#3;#4;#5 {{\frenchspacing\par\rn#1 #2, {\it #3} (#4: #5)\par}}
\def\rfproc#1;#2;#3;#4;#5;#6 {{\frenchspacing\par\rn#1 #2, in {\it #3}, ed. #4 (#5: #6)\par}}
\def\rfprocp#1;#2;#3;#4;#5;#6;#7 {{\frenchspacing\par\rn#1 #2, in {\it #3}, ed. #4 (#5: #6), p#7\par}}
\def\rfprep#1;#2;#3  {{\par\rn#1 #2, #3\par}}
\def\rfprepp#1;#2;#3 {{\par\rn#1 #2, #3\par}}

\def\expec#1{\langle#1\rangle}

\def\etal{{\frenchspacing\it et al.}}
\def\ie{{\frenchspacing\it i.e.}}
\def\eg{{\frenchspacing\it e.g.}}
\def\etc{{\frenchspacing\it etc.}}

\def\beq#1{\begin{equation}\label{#1}}
\def\eeq{\end{equation}}
\def\beqa#1{\begin{eqnarray}\label{#1}}
\def\eeqa{\end{eqnarray}}
\def\eq#1{equation~(\ref{#1})}
\def\Eq#1{Equation~(\ref{#1})}
\def\eqn#1{~(\ref{#1})}

\def\fig#1{Figure~\ref{#1}}
\def\Fig#1{Figure~\ref{#1}}

\def\sec#1{Section~\ref{#1}}

\def\spose#1{\hbox to 0pt{#1\hss}}
\def\simlt{\mathrel{\spose{\lower 3pt\hbox{$\mathchar"218$}}
     \raise 2.0pt\hbox{$\mathchar"13C$}}}
\def\simgt{\mathrel{\spose{\lower 3pt\hbox{$\mathchar"218$}}
     \raise 2.0pt\hbox{$\mathchar"13E$}}}
\def\simpropto{\mathrel{\spose{\lower 3pt\hbox{$\mathchar"218$}}
     \raise 2.0pt\hbox{$\propto$}}}

\def\ed{\end{document}}

\def\fsky{f_{\rm sky}}

\def\Ol{\Omega_\Lambda}
\def\Om{\Omega_m}

\def\expec#1{\langle#1\rangle}

\def\etal{{\frenchspacing\it et al.}}
\def\ie{{\frenchspacing\it i.e.}}
\def\eg{{\frenchspacing\it e.g.}}
\def\etc{{\frenchspacing\it etc.}}

\def\beq#1{\begin{equation}\label{#1}}
\def\eeq{\end{equation}}
\def\beqa#1{\begin{eqnarray}\label{#1}}
\def\eeqa{\end{eqnarray}}
\def\eq#1{equation~(\ref{#1})}
\def\Eq#1{Equation~(\ref{#1})}
\def\eqn#1{~(\ref{#1})}

\newcommand{\beeq}{\begin{equation}} 
\newcommand{\beeqa}{\begin{eqnarray}}

\def\fig#1{Figure~\ref{#1}}
\def\Fig#1{Figure~\ref{#1}}

\def\sec#1{Section~\ref{#1}}

\def\spose#1{\hbox to 0pt{#1\hss}}
\def\simlt{\mathrel{\spose{\lower 3pt\hbox{$\mathchar"218$}}
     \raise 2.0pt\hbox{$\mathchar"13C$}}}
\def\simgt{\mathrel{\spose{\lower 3pt\hbox{$\mathchar"218$}}
     \raise 2.0pt\hbox{$\mathchar"13E$}}}
\def\simpropto{\mathrel{\spose{\lower 3pt\hbox{$\mathchar"218$}}
     \raise 2.0pt\hbox{$\propto$}}}

\def\ed{\end{document}}

\def\tr{\hbox{tr}\>}
\def\ln{\,\hbox{ln}\,}

\def\nbar{{\bar n}}
\def\Nbar{{N}}

\def\ith{i^{\rm th}}
\def\b{{\bf b}}
\def\e{{\bf e}}

\def\p{{\bf p}}
\def\phat{\widehat\p}
\def\ph{\widehat p}

\def\x{{\bf x}}

\def\y{{\bf y}}

\def\A{{\bf A}}
\def\B{{\bf B}}
\def\C{{\bf C}}
\def\D{{\bf D}}

\def\F{{\bf F}}
\def\I{{\bf I}}

\def\L{{\bf L}}
\def\M{{\bf M}}
\def\N{{\bf N}}
\def\P{{\bf P}}

\def\W{{\bf W}}

\def\Q{{\bf Q}}

\def\S{{\bf S}}

\def\aa{\alpha}
\def\bb{\beta}
\def\cc{\gamma}
\def\zero{{\bf 0}}

\def\M{{\bf M}}

\def\tr{\hbox{tr}\>}

\def\l{\ell}

\def\ed{\end{document}}

\def\tr{\hbox{tr}\,}

\def\ith{i^{th}}

\documentstyle[emulateapj,danonecolfloat]{article}

\begin{document}
\twocolumn[


\journalid{337}{15 January 1989}
\articleid{11}{14}

\def\penn{1}
\def\chicago{2}
\def\fnal{3}
\def\tucson{4}
\def\pton{5}
\def\nyu{6}
\def\ias{7}
\def\pitt{8}
\def\columbia{9}
\def\hopkins{10}
\def\sussex{11}
\def\cmu{12}
\def\hawaii{13}
\def\drexel{14}
\def\apo{15}
\def\tokyo{16}
\def\flagstaff{17}
\def\michigan{18}
\def\rochester{19}
\def\efi{20}

\submitted{Submitted to ApJ July 2 2001, accepted January 15 2002}

\title{The Angular Power Spectrum of Galaxies from Early SDSS Data}

\author{Max Tegmark\altaffilmark{\penn}, 
Scott Dodelson\altaffilmark{\chicago,\fnal},
Daniel J. Eisenstein\altaffilmark{\tucson},
Vijay Narayanan\altaffilmark{\pton},
Roman Scoccimarro\altaffilmark{\nyu,\ias}, 
Ryan Scranton\altaffilmark{\chicago,\fnal},
Michael A. Strauss\altaffilmark{\pton}, 
Andrew Connolly\altaffilmark{\pitt},
Joshua A. Frieman\altaffilmark{\chicago,\fnal},
James E. Gunn\altaffilmark{\pton}, 
Lam Hui\altaffilmark{\columbia}, 
Bhuvnesh Jain\altaffilmark{\penn},
David Johnston\altaffilmark{\chicago,\fnal}, 
Stephen Kent\altaffilmark{\fnal},
Jon Loveday\altaffilmark{\sussex}, 
Robert C. Nichol\altaffilmark{\cmu}, 
Liam O'Connell\altaffilmark{\sussex},
Ravi K. Sheth\altaffilmark{\fnal}, 
Albert Stebbins\altaffilmark{\fnal},
Alexander S. Szalay\altaffilmark{\hopkins},
Istv\'an Szapudi\altaffilmark{\hawaii}, 
Michael S. Vogeley\altaffilmark{\drexel},
Idit Zehavi\altaffilmark{\fnal}, 
James Annis\altaffilmark{\fnal}, 
Neta A. Bahcall\altaffilmark{\pton}, 
J. Brinkmann\altaffilmark{\apo},
Istvan Csabai\altaffilmark{\hopkins},
Mamoru Doi\altaffilmark{\tokyo},  
Masataka Fukugita\altaffilmark{\tokyo},
Greg Hennessy\altaffilmark{\flagstaff},
\v Zeljko Ivez\'ic\altaffilmark{\pton},
Gillian R. Knapp\altaffilmark{\pton},
Don Q. Lamb\altaffilmark{\chicago},
Brian C. Lee\altaffilmark{\fnal},
Robert H. Lupton\altaffilmark{\pton},
Timothy A. McKay\altaffilmark{\michigan},
Peter Kunszt\altaffilmark{\hopkins},
Jeffrey A. Munn\altaffilmark{\flagstaff}, 
John Peoples\altaffilmark{\fnal}, 
Jeffrey R. Pier\altaffilmark{\flagstaff},
Michael Richmond\altaffilmark{\rochester},
Constance Rockosi\altaffilmark{\chicago}, 
David Schlegel\altaffilmark{\pton}, 
Christopher Stoughton\altaffilmark{\fnal}, 
Douglas L. Tucker\altaffilmark{\fnal},
Brian Yanny\altaffilmark{\fnal}, 
Donald G. York\altaffilmark{\chicago,\efi},
for the SDSS Collaboration
}
\footnote{\penn}{Department of Physics, University of Pennsylvania,
Philadelphia, PA 19101, USA}
\footnote{\chicago}{Astronomy and Astrophysics Department, University of
Chicago, Chicago, IL 60637, USA}
\footnote{\fnal}{Fermi National Accelerator Laboratory, P.O. Box 500, Batavia,
IL 60510, USA}
\footnote{\tucson}{Department of Astronomy, University of Arizona, 
Tucson, AZ 85721, USA}
\footnote{\pton}{Princeton University Observatory, Princeton, NJ 08544,
USA}
\footnote{\nyu}{Department of Physics, New York University, 4 Washington
Place, New York, NY 10003}
\footnote{\pitt}{University of Pittsburgh, Department of Physics and
Astronomy, 3941 O'Hara Street, Pittsburgh, PA 15260, USA}
\footnote{\columbia}{Department of Physics, Columbia University, New York, NY
10027, USA}
\footnote{\hopkins}{Department of Physics and Astronomy, The Johns Hopkins
University, 3701 San Martin Drive, Baltimore, MD 21218, USA}
\footnote{\sussex}{Sussex Astronomy Centre, University of Sussex, Falmer,
Brighton BN1 9QJ, UK}
\footnote{\cmu}{Department of Physics, 5000 Forbes Avenue, Carnegie
Mellon
University, Pittsburgh, PA 15213, USA}
\footnote{\ias}{Institute for Advanced Study, School of Natural
Sciences,
Olden Lane, Princeton, NJ 08540, USA}
\footnote{\hawaii}{Institute for Astronomy, University of Hawaii, 2680
Woodlawn Drive, Honolulu, HI 96822, USA}
\footnote{\drexel}{Department of Physics, Drexel University, Philadelphia,
PA
19104, USA}
\footnote{\apo}{Apache Point Observatory, 2001 Apache Point Rd, 
Sunspot, NM 88349-0059, USA}
\footnote{\tokyo}{Inst. for Cosmic Ray Research, 
Univ. of Tokyo, Kashiwa 277-8582, Japan}
\footnote{\flagstaff}{U.S. Naval Observatory, 
Flagstaff Station, Flagstaff, AZ 86002-1149, USA}
\footnote{\michigan}{Dept. of Physics, Univ. of Michigan, 
Ann Arbor, MI 48109-1120, USA}
\footnote{\rochester}{Physics Dept., Rochester Inst. of Technology, 
1 Lomb Memorial Dr, Rochester, NY 14623, USA}
\footnote{\efi}{Enrico Fermi Institute, University of
Chicago, Chicago, IL 60637, USA}

\begin{abstract}
We compute the angular power spectrum $C_\l$ from
1.5 million galaxies in early SDSS data
on large angular scales, $\l\simlt 600$.
The data set covers about 160 square degrees, with a characteristic
depth of order 
$1 h^{-1}$ Gpc 
in the faintest ($21<r^*<22$) of our four
magnitude bins.
Cosmological interpretations of these results are
presented in a companion paper by Dodelson {\etal} (2001).
The data in all four magnitude bins are consistent with a
simple flat ``concordance'' model with nonlinear evolution
and linear bias factors of order unity. Nonlinear evolution
is particularly evident for the brightest galaxies.
A series of tests suggest that systematic errors related to
seeing, reddening, etc., are negligible, 
which bodes well for the sixtyfold larger
sample that the SDSS is currently collecting.
Uncorrelated error bars and well-behaved window functions
make our measurements a convenient starting point for
cosmological model fitting.
\end{abstract}

]


\section{INTRODUCTION}

Galaxy clustering encodes a wealth of cosmological information.
By breaking degeneracies between cosmological parameters
and by permitting powerful cross checks, it 
complements other cosmological probes
such as the cosmic microwave background (CMB) both 
in theory (\eg, Eisenstein {\etal} 1999)
and in practice 
(\eg, Netterfield {\etal} 2001; Pryke {\etal} 2001;
Stompor {\etal} 2001; Wang {\etal} 2001).

Although purely angular galaxy catalogs lack the 
three-dimensional (3D) information present in redshift surveys,
they tend to be quite competitive because of their 
much greater numbers of galaxies. A case in point
is the APM survey, which still provides one of 
the most accurate three-dimensional power spectrum measurements despite
lacking redshift information
(Efstathiou \& Moody 2001).
In this spirit, the goal of the present paper is to 
measure the two-dimensional (2D) power spectrum $C_\l$ 
from early imaging data in the Sloan Digital
Sky Survey
(SDSS; York {\etal} 2000).
The angular correlation function $w(\theta)$ of this SDSS data 
is presented in a companion paper by Connolly {\etal}
(2001), and both of these angular clustering 
measures are inverted to 3D power spectra 
$P(k)$ by Dodelson {\etal} (2001). 
The galaxies are analyzed directly in terms of power
spectrum parameters by Szalay {\etal} (2001). 
The data set upon which all these analyses
are based is presented and extensively tested for systematic
errors by Scranton {\etal} (2001, hereafter S2001).
3D clustering using galaxies with measured redshifts is 
studied by Zehavi {\etal} (2001). An independent $w(\theta)$-analysis
is presented by Gazta\~naga (2001).

The angular correlation function $w(\theta)$ has many merits
as a measure of clustering.
It is fast to compute even for massive data sets, 
and its broad familiarity in the astronomical community 
facilitates comparison with theoretical predictions as
well as other observations.
Notwithstanding, as detailed in Appendix A, 
the angular power spectrum $C_\l$ 
has three virtues that makes it quite complementary to 
$w(\theta)$ and worth computing as well\footnote{
It is worth emphasizing that although 
the {\it theoretical} $C_\l$ and $w(\theta)$ are
simply Fourier (more precisely Legendre) transforms
of one another, there is no such equivalence between
the {\it measured} $C_\l$ and $w(\theta)$ 
because of incomplete sky coverage and other complications.
Because different pair weightings are applied to the
multitude of galaxies before they are compressed into
the handful of $C_\l$ and $w(\theta)$ numbers presented here
and by Connolly {\etal} (2001), the information content in the two is different. 
Although it is possible to construct a lossless
$w(\theta)$-estimator that contains the same information as $C_\l$,
this is not desirable for the reasons described in Appendix A ---
it limits the dynamic range and it destroys a
key property of conventional $w(\theta)$-estimators:
perfect window functions, \ie, the estimated 
correlation at separation $\theta$ probes only 
correlations on that scale.
}:
\begin{enumerate}
\item 
It is possible to produce measurements of 
$C_\l$ that have both uncorrelated errors and 
well-behaved window functions.
\item
The $C_\l$-estimators represent a lossless 
compression of the full data set in 
the sense that they retain all of its angular
clustering information on large scales, where the 
Gaussian approximation applies.
\item The $C_\l$-coefficients are 
more closely related to the 3D power spectrum $P(k)$
than $w(\theta)$ is, in the sense of giving narrower window
functions in $k$-space (Baugh \& Efstathiou 1994). This is an advantage 
for 2D $\mapsto$ 3D inversions, since it reduces troublesome
aliasing from small scales where nonlinear effects are 
difficult to model.
\end{enumerate}
These attractive properties have triggered a resurgence of interest in 
measuring $C_\l$ from galaxy surveys (Scharf \& Lahav 1993;
Baugh \& Efstathiou 1994; Huterer {\etal} 2000), extending the
pioneering work of Hauser \& Peebles (1973).

On small scales where nonlinear effects become important, 
the angular power spectrum loses much of its appeal.
Non-Gaussian clustering introduces correlations
between different $\l$-bands, our method becomes computationally
cumbersome, and much of the interesting
physics takes place in real space rather than in 
Fourier space, with the observed clustering telling us
more about halo properties than about the initial
linear power spectrum. 
In summary, as described in Appendix A, the 
$C_\l$-analysis presented here and the $w(\theta)$ analysis
by Connolly {\etal} (2001) are highly complementary, with
advantages on large and small scales, respectively.
We therefore limit our analysis 
to large angular scales $\l\simlt 600$, corresponding
to the linear and weakly nonlinear regime. A multipole
$\l$ corresponds roughly to an angular
scale $\theta\sim 180^\circ/\ell$, so our limit
$\l\simlt 600$ corresponds to a spatial scale of 
order $5h^{-1}$ Mpc at the characteristic survey 
depth of $1h^{-1}$ Gpc.

The rest of this paper is organized as follows.
In \sec{ClSec}, we measure the angular power spectrum $C_\l$
and discuss how it is related to the underlying 3D power
spectrum $P(k)$. 
In \sec{SystematicsSec}, we perform a range of tests and 
Monte-Carlo studies to assess
the reliability of our results given potential problems
with extinction, seeing, software and non-linear
clustering, and summarize our conclusions.
Two appendices discuss how our 
angular power spectrum measurements relate to the angular correlation
function $w(\theta)$ and the underlying 3D power spectrum $P(k)$.

\section{The angular power spectrum}
\label{ClSec}

\subsection{Data}

This paper builds on the foundation laid by S2001, which 
produces a galaxy sample demonstrated to be of sufficient
quality to permit a large-scale angular clustering analysis
not dominated by systematic errors.
We use the ``EDR-P'' sample described by S2001 
for our analysis, which stands for 
{\bf e}arly {\bf d}ata {\bf r}elease 
(Stoughton {\etal} 2001)
with galaxy {\bf p}robabilities 
used in place of rigid counts\footnote{As detailed by S2001,
each object is assigned a probability between zero and one
that it is a galaxy based on its observed properties.
Throughout this paper, we use the sum of these probabilities
as our estimate of the number of galaxies in a given region.
This is more accurate than a strict object-by-object 
maximum-likelihood classification --- for instance, 
if ten objects each have a 10\% probability of being a galaxy, 
classifying them all as stars would underestimate the 
true galaxy count by one. 
}.
It consists of galaxies in the $2.5^\circ\times 90^\circ$ equatorial 
stripe $145^\circ <\alpha_{2000} < 235^\circ$, 
$-1.25^\circ<\delta_{2000}<1.25^\circ$
with regions of high extinction and poor seeing discarded.
We measured fluxes with the $r$ filter.
The $r$ magnitude is defined by Fukugita {\etal} (1996), 
Stoughton {\etal} (2001).
As in S2001, we analyze four subsamples of the galaxies separately,
corresponding to ranges of model magnitude $r^*$ of 
18-19, 19-20, 20-21 and 21-22, respectively.
These four samples consist of effectively 
$N=$57,781,
158,636,
428,920 and 
886,936 galaxies, respectively,
with assumed mean redshifts of 0.26, 0.36, 0.50 and 0.64,
respectively. Assuming a flat $\Omega_\Lambda=0.7$ cosmology,
this corresponds to mean comoving distances of 
0.51,   0.71, 0. 95 and 1.19 $h^{-1}$Gpc, respectively.

A set of powerful tools for angular power spectrum estimation
has been developed in the CMB community, and to take advantage
of this, we begin by re-expressing our galaxy analysis problem in
a form analogous to the CMB case.
We do this by dividing our sky patch into $N$ square 
``pixels'' of side 12.5 arcminutes
and computing the density fluctuation
\beq{xDefEq}
x_i\equiv{n_i\over\nbar_i} - 1
\eeq
in each one. Here $n_i$ is the observed 
number of galaxies in each pixel and $\nbar_i$ is the
expected number, taking into account the slight
spatial variations in completeness  as in S2001.
The choice of 12.5' for the pixel height is convenient since
it correspoonds the height of an SDSS camera column
(Gunn {\etal} 1998), thereby maximizing the sensitivity
of our tests for weather-related systematics (even and odd columns are
observed on separate occasions).
There are 3695 pixels in each of the three brightest 
magnitude bins and 3274 in the $21<r^*<22$ bin where the seeing cuts were
more stringent, corresponding to sky areas of 
160 and 142 square degrees, respectively.

In the context of previous large angular surveys of galaxies,
the main advantage of our data set is its superior photometric accuracy.
Its main drawback is that it subtends less area than 
both the APM and EDSGC surveys, which covered 5000 
and 1000 square degrees, respectively
 (see Efstathiou \& Moody 2001;  Huterer {\etal} 2000).
This weakess is partly compensated by going deeper
(our sample of 1.5 million galaxies is about half that
of APM and 50\% larger than that of EDSGC) and is of course
only temporary, since the SDSS will ultimately cover $10^4$ square degrees.

%

\subsection{The basic problem}

Given a pixelized map $x_i$ and associated shot noise 
error bars $\nbar_i^{-1/2}$,
we compute the angular power spectrum with the quadratic
estimator method (Tegmark 1997; Bond {\etal} 2000), 
using KL-compression to accelerate the process
(Bond 1994; Bunn 1995; Vogeley \& Szalay 1996).
Since this procedure has been described in detail in the recent
literature 
(see Tegmark \& de Oliveira-Costa 2001 for a recent review
using our present notation and Huterer {\etal} (2001) for a recent 
application to galaxy clustering),
we summarize the method only very briefly here.

We group our angular density fluctuation map pixels $x_i$ into an $N$-dimensional vector $\x$.
The vector $\x$ has a vanishing expectation value 
(${\expec{\x}=\zero}$) by construction, and we can write its covariance matrix
as
\beq{CdefEq}
\C\equiv \expec{\x\x^t}= \S+\N,\quad \S\equiv\sum_i p_i \P_i,
\eeq
for a set of angular power spectrum parameters $p_i$ and known matrices $\P_i$
that are given by the map geometry in terms of 
Legendre polynomials.
$\N$ denotes the contribution from shot noise, and is 
a known diagonal matrix.
We parametrize the angular power spectrum
\beq{sigmaDefEq}
\sigma_\l^2 \equiv {\l(\l+1)\over 2\pi}C_\l
\eeq
(customarily denoted $\delta T^2_\l$ in the CMB literature)
as piecewise constant in 50 bands of width $\Delta\l=20$,
with height $p_i$ in the $\ith$ band.
$\sigma_\l$, which is a dimensionless number, 
can roughly be interpreted as the rms fluctuation level on the
angular scale $\theta\sim 180^\circ/\ell$.
In summary, knowing the power spectrum parameters $p_i$ would allow us to
predict the theoretical covariance matrix of our data via 
\eq{CdefEq}. Our problem is to do the opposite, and 
estimate the parameters $p_i$ using the observed data vector $\x$.

\subsection{KL-compression}

\begin{figure}[tb] 
\centerline{\epsfxsize=\figsize\epsffile{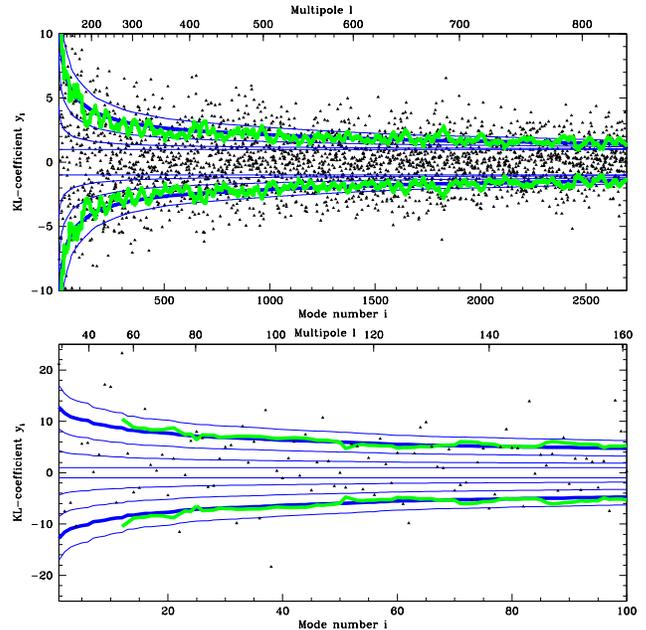}}
\vskip-0.5cm
\caption{\label{rmsFig}\footnotesize%
The triangles show the KL-coefficients 
$y_i$ for the $21<r^*<22$ magnitude bin (2693 in top panel, 
first 100 in bottom panel).
If there were no clustering in the survey, merely shot noise, 
they would have unit variance, and about $68\%$ of them would be
expected to lie 
between the two horizontal lines (in the band
$-1<y_i<1$).
Cosmological fluctuations in the data would
increase the standard deviation, as indicated by the 
other thin curves. 
From inside out, the thin curves correspond to 
the theoretically predicted rms fluctuation level
$\sqrt{1+\lambda_i}$ with
our prior power spectrum renormalized by factors
0, $(1/3)^2$, $(2/3)^2$, 1 and $(4/3)^2$, respectively.
The green/grey curve is the rms of the data points
$y_i$, averaged in bands of width 25, and shows the cosmological 
fluctuation signal rising to the left.
As a rule of thumb, the $\ith$ KL-mode probes angular scales
$\l\sim (i/\fsky)^{1/2}\sim 16\sqrt{i}$, as indicated by upper axis labels,
where
$\fsky\approx 0.004$ is the sky fraction covered in
this survey (Tegmark 1997).
Note that green/grey curves nearly match the power per mode
predicted by the prior model curve (normalization factor 1, heavy curve), 
showing that this model is good for estimating errors.
}
\end{figure}

Since the power spectrum estimation in the next subsection involves
repeatedly multiplying and inverting $N\times N$ matrices,
and each such manipulation requires of order $N^3$ operations, 
we apply a data-compression step that reduces the size of our data set.
We employ the
Karhunen-Lo\`eve (KL) compression method
(Karhunen 1947; Bond 1995; Bunn \& Sugiyama 1995; Vogeley \& Szalay 1996;
Tegmark {\etal} 1997; Szalay {\etal} 2001), 
which compresses 
the information content of a map into the first part of 
a vector $\y\equiv \B^t\x$, 
where $\B$ is an $N\times N$ matrix whose $i^{th}$ column $\b_i$ 
satisfies the generalized eigenvalue equation
\beq{EigenEq}
\S\b_i = \lambda_i\N\b_i,
\eeq
normalized so that $\b^t_i\N\b_i=1$ and sorted by decreasing $\lambda_i$.
The $N$ numbers $y_i$ are uncorrelated, \ie, 
\beq{yCovEq}
\expec{y_i y_j} = \b^t_i (\N + \S) \b_j = (1+\lambda_i)\delta_{ij},
\eeq
and their variance $\expec{y_i^2}$ has a contribution of $1$ from noise and 
$\lambda_i$ from signal. This means that the eigenvalue $\lambda_i$ can be 
interpreted as a signal-to-noise ratio for $y_i$.
The first 500 of these numbers $y_i$ (KL-coefficients) are shown in 
\fig{rmsFig} for the $21<r^*<22$ band, and 
it is seen that most of the cosmological signal is contained in
the first few hundred modes. We discard all modes with signal-to-noise
ratio $\lambda_i$ below unity, which leaves us with 
1255, 1656, 2510 and 2693 modes for the four magnitude bands, respectively.
This KL-expansion is useful not only to save time, but also for systematic
error checks. \Fig{rmsFig} shows that none of the modes deviates
from zero by a surprisingly large amount (for instance, 
out of the first 100 modes, typically only 5 should deviate by 2$\sigma$ 
and none by $3\sigma$).
A similar KL-compression is performed in Szalay {\etal} (2001),
where parameters of the 3D power spectrum are measured directly from
the KL modes. 2D images of KL-modes for a rectangular strip 
are plotted by Tegmark (1997) and Szalay {\etal} (2001), illustrating
that they tend to probe progressively smaller angular scales.

\subsection{Integral constraint}

An important complication when computing clustering on large scales is the
so-called integral constraint. Since the mean galaxy density $\nbar$ 
is a priori unknown, it must be estimated from the data itself,
implicitly forcing the vector $\x$ to have zero mean.
We tackle this problem by only using modes that are orthogonal
to the (completely unknown) mean, \ie, to the vector
$\e=(1,1,...,1)$ corresponding to a constant offset in the map.
This idea goes back to Fisher {\etal} (1993) and becomes 
very simple to implement for our pixelized case (Tegmark {\etal} 1998).
In principle, it suffices to add a very large noise to the mean mode,
\ie, to add a huge number $M$ times $\e\e^t$ to the noise 
matrix $\N$, and the subsequent KL-compression will automatically 
relegate the mean mode to the list of useless ones to be discarded.
In practice, we remove the mean mode analytically as described
in Appendix B of Tegmark {\etal} (1998), which corresponds to the
limit where the huge number $M\to\infty$.

\subsection{Basic results}
\label{ResultsSec}

\begin{figure}[tb] 
\vskip-1.0cm
\centerline{\epsfxsize=\figsize\epsffile{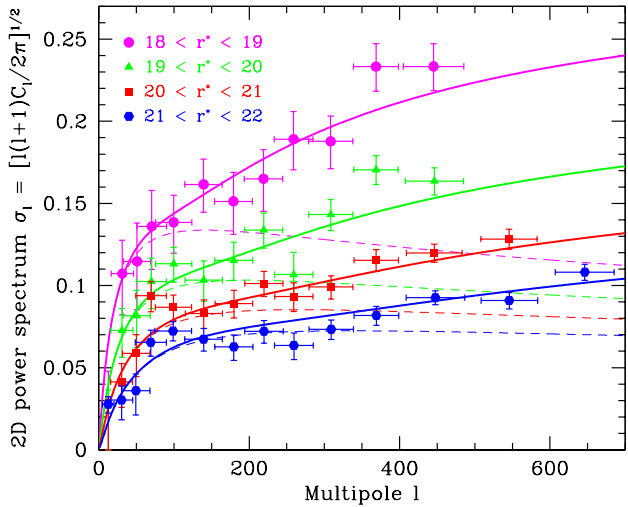}}
\vskip-1.0cm
\caption{\label{ClFig}\footnotesize%
The angular power spectrum $\sigma_\l\equiv[\l(\l+1)C_\l /2\pi]^{1/2}$ 
is shown for the four magnitude bins.
The horizontal location of each point and the
associated horizontal bars 
correspond to the mean and rms width of the corresponding
window function. 
These measurements are uncorrelated in the approximation
of Gaussian fluctuations.
The curves are the ``prior''
power spectra used, \protect\ie, 
the concordance model from Wang {\protect\etal} (2001) with (solid)
and without (dashed) nonlinear evolution, using 
four separate bias factors of order unity as 
described in the text.
}
\end{figure}

\begin{figure}[tb] 
\vskip-1.0cm
\centerline{\epsfxsize=\figsize\epsffile{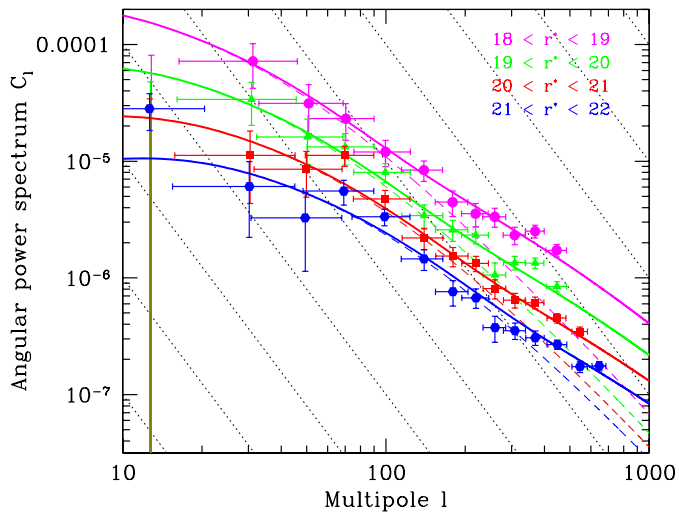}}
\vskip-1.0cm
\caption{\label{ClFig2}\footnotesize%
Same as previous figure, but with logarithmic axes
and for $C_\l$ rather than $\sigma_\l$.
Although logarithmic axes make window functions more difficult 
to interpret, it facilitates connecting to the 
underlying 3D power spectrum $P(k)$, which is 
very crudely speaking the same curve shifted vertically
and horizontally with different axis labels. 
The shifts depend on the magnitude bin:
the fainter (and on average more distant) the galaxies, the 
further up and to the left the curve should be shifted
--- up because there is more averaging along the line of sight
which suppresses fluctuations, to the left because
a given angular scale $\l$ corresponds to larger 
spatial scales. The solid lines of slope $-3$ indicate the
direction of this shift when the mean survey depth is changed.
In the absence of relative bias, this shifting should place
the four curves on top of each other.
}
\end{figure}

Once our data and the corresponding matrices have been KL-compressed
(in which $\x$ gets replaced by $\y\equiv{\B'}^t\x$, 
$\P_i$ gets replaced by ${\B'}^t\P_i\B'$, 
$\N$ gets replaced by ${\B'}^t\N\B'=\I$, where the rectangular matrix 
$\B'$ denotes the left part of the square matrix $\B$ corresponding to 
the KL column vectors we wish to keep),
we proceed to compute
quadratic estimators $\ph_i$ of 
our power spectrum parameters $p_i$.
The results are shown in figures~\ref{ClFig} and~\ref{ClFig2} 
and are listed in Table 1.

Since it is important for the interpretation, let us briefly review
how these measurements are computed from the input
data, in this case the vector $\y$ of KL-modes.
A quadratic estimator $\ph_i$ is simply a quadratic 
function of the data vector,
so the most general unbiased case can be written as
\beq{qDefEq}
\ph_i\equiv\y^t\Q_i\y - s_i,
\eeq
where the $\Q_i$ are arbitrary symmetric $N\times N$-dimensional matrices
and the $s_i\equiv\tr[\Q_i\N]$ are the shot noise contributions.
Grouping the parameters 
$p_i$ and the estimators $\ph_i$ into  
vectors denoted $\p$ and $\phat$, the expected measurement is
\beq{qMeanEq}
\expec{\phat}=\W\p
\eeq
for a {\it window matrix} $\W$ that can be computed 
from the $\Q_i$-matrices and the sky geometry alone
($\W_{ij}=\tr[\P_i\Q_j])$. 
The $\Q$-matrices are normalized so that each row of the
window matrix sums to unity. This enables us to interpret
each band power measurement $\ph_i$ as a weighted average of
the true power spectrum $p_j$, the elements of
the $\ith$ row of $\W$ giving the weights (the ``window function'').

The basic idea with quadratic estimators is that each matrix 
$\Q_i$ can be chosen to 
effectively Fourier transform the sky map, square the Fourier modes
in the $\ith$ power spectrum band and average the results together,
thereby probing the power spectrum on that scale.
We use the particular choice of $\Q$-matrices advocated by
Tegmark \& Hamilton (1998) (see Tegmark \& Oliveira-Costa 2001
for a treatment conforming to our notation), described in Appendix A,
which has the advantage of making the error bars on the
measurements uncorrelated. In other words, 
the covariance matrix for the measured vector $\phat$ is diagonal
(combining shot noise and sample variance errors), so it is completely
characterized by its diagonal elements, given by the error bars in 
Table 1 and \fig{ClFig}. 
This covariance matrix 
$\M\equiv\expec{\phat\phat^t}-\expec{\phat}\expec{\phat}^t$
is generally given by 
$\M_{ij}= 2\tr[\Q_i\C\Q_j\C]$
for the Gaussian case, and our particular choice of 
$\Q$-matrices thus reduces it to a diagonal matrix
$\M_{ij}=\delta_{ij}(\Delta\ph_i)^2$. $\C$ of course depends on $\p$
through \eq{CdefEq}, and when computing $\M$ to obtain our
error bars $\Delta\ph_i$, we use the ``prior'' power spectra described
below, smooth curves fitting our measurements.

\begin{figure}[tb] 
\vskip-1.0cm
\centerline{\epsfxsize=\figsize\epsffile{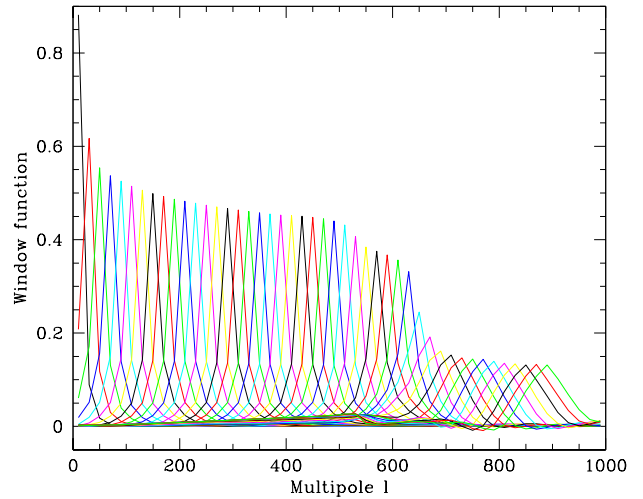}}
\vskip-1.0cm
\caption{\label{lwindowFig}\footnotesize%
Sample window functions
are shown for the band-power measurements in the
magnitude bin $21<r^*<22$. These are the rows
of the window matrix $\W$, and connect our band-power
measurements $\ph_i$ to the underlying power spectrum $\sigma_\l^2$.
}
\end{figure}

The window functions corresponding to our
50 band power measurements (the rows of the matrix $\W$)
are plotted in \fig{lwindowFig} for the faintest 
magnitude bin.
This connects our measurements $\ph_i$ to the binned underlying
power spectrum $\sigma_\l^2$.
The windows are seen to have a characteristic 
width of order $\Delta\l\sim 50$, which is determined
by the size of our sky patch in the narrowest direction
(Tegmark 1997).
We are thus unable to 
resolve the angular power spectrum finer than this because our
survey strip is so narrow in the declination direction, 
limiting the $\l$-resolution to of order $\Delta\l\sim 180^\circ/2.5^\circ$.
\Fig{lwindowFig} also shows a notable transition around 
$\l=600$. This coincides with the angular scale 
where the cosmological fluctuations drop below Poissonian 
shot noise fluctuations, and has a simple interpretation.
On the larger scales where shot noise is less important, the
$\Q$-matrices weight the galaxies in such a way as to make the
window functions narrow, thereby minimizing the 
sample variance contribution to the error bars
caused by power aliased from other scales.
On smaller scales,
the $\Q$-matrices weight all areas of the map essentially equally,
without bothering with niceties such as apodization (down-weighting 
parts near edges), in an attempt to minimize the all-dominating shot noise.
This results in less well-behaved window functions, which are both
broader and are seen to have a ``red leak'' of power from substantially
larger scales. Since the measurements beyond this transition regime are
noise dominated and 
contain very little information, producing mere upper limits, 
we simply discard them. This cutoff corresponds to 
$\l=$500, 500, 600 and 700 in the four magnitude bins, respectively
--- note that shot noise
dominates the brighter magnitude bins at lower $\l$, since
they contain fewer galaxies.

To improve the signal-to-noise ratio, we average these measurements 
into bands as specified in Table 1. Since the original
measurements are uncorrelated, so are these averages.
The corresponding $14\times 50$ window function matrices
for each magnitude bin, 
which are necessary for comparing our
measurements with theoretical predictions, will be published
electronically with this article and are also available 
at {\it http://www.hep.upenn.edu/$\sim$max/sdss.html}.

\begin{table*}[h]
\begin{center}
{\small
\begin{tabular}{|cc|cc|cc|cc|} 
\hline 
\multicolumn{2}{|c}{$18<r<19$}&
\multicolumn{2}{|c}{$19<r<20$}&
\multicolumn{2}{|c}{$20<r<21$}&
\multicolumn{2}{|c|}{$21<r<22$}\\
$\l$&$\sigma_\l^2$&
$\l$&$\sigma_\l^2$&
$\l$&$\sigma_\l^2$&
$\l$&$\sigma_\l^2$\\
\hline 
13 $\pm$ 8 & -0.0006 $\pm$ 0.0028 & 13 $\pm$ 8 & 0.0002 $\pm$ 0.0011 & 13 $\pm$ 8 & 0.0004 $\pm$ 0.0005 & 13 $\pm$ 8 & 0.0008 $\pm$ 0.0003\\
31 $\pm$ 15 & 0.0115 $\pm$ 0.0048 & 31 $\pm$ 15 & 0.0053 $\pm$ 0.0021 & 30 $\pm$ 15 & 0.0017 $\pm$ 0.0010 & 30 $\pm$ 15 & 0.0009 $\pm$ 0.0006\\
51 $\pm$ 18 & 0.0132 $\pm$ 0.0059 & 50 $\pm$ 18 & 0.0067 $\pm$ 0.0028 & 50 $\pm$ 18 & 0.0034 $\pm$ 0.0015 & 49 $\pm$ 19 & 0.0013 $\pm$ 0.0008\\
70 $\pm$ 20 & 0.0185 $\pm$ 0.0064 & 70 $\pm$ 20 & 0.0105 $\pm$ 0.0031 & 70 $\pm$ 20 & 0.0088 $\pm$ 0.0017 & 69 $\pm$ 21 & 0.0043 $\pm$ 0.0010\\
100 $\pm$ 24 & 0.0192 $\pm$ 0.0048 & 100 $\pm$ 24 & 0.0128 $\pm$ 0.0024 & 99 $\pm$ 24 & 0.0075 $\pm$ 0.0013 & 99 $\pm$ 24 & 0.0052 $\pm$ 0.0008\\
139 $\pm$ 25 & 0.0261 $\pm$ 0.0052 & 140 $\pm$ 25 & 0.0107 $\pm$ 0.0025 & 140 $\pm$ 25 & 0.0069 $\pm$ 0.0014 & 139 $\pm$ 25 & 0.0045 $\pm$ 0.0009\\
179 $\pm$ 25 & 0.0229 $\pm$ 0.0056 & 179 $\pm$ 25 & 0.0133 $\pm$ 0.0027 & 180 $\pm$ 25 & 0.0079 $\pm$ 0.0015 & 180 $\pm$ 25 & 0.0039 $\pm$ 0.0010\\
219 $\pm$ 26 & 0.0272 $\pm$ 0.0061 & 219 $\pm$ 25 & 0.0179 $\pm$ 0.0028 & 220 $\pm$ 25 & 0.0103 $\pm$ 0.0015 & 220 $\pm$ 26 & 0.0052 $\pm$ 0.0010\\
259 $\pm$ 26 & 0.0357 $\pm$ 0.0067 & 259 $\pm$ 26 & 0.0114 $\pm$ 0.0030 & 259 $\pm$ 25 & 0.0087 $\pm$ 0.0016 & 260 $\pm$ 26 & 0.0040 $\pm$ 0.0010\\
308 $\pm$ 29 & 0.0353 $\pm$ 0.0060 & 309 $\pm$ 29 & 0.0205 $\pm$ 0.0027 & 309 $\pm$ 29 & 0.0098 $\pm$ 0.0014 & 309 $\pm$ 29 & 0.0054 $\pm$ 0.0009\\
369 $\pm$ 30 & 0.0544 $\pm$ 0.0067 & 369 $\pm$ 30 & 0.0291 $\pm$ 0.0030 & 369 $\pm$ 29 & 0.0133 $\pm$ 0.0015 & 369 $\pm$ 30 & 0.0067 $\pm$ 0.0009\\
445 $\pm$ 40 & 0.0545 $\pm$ 0.0066 & 446 $\pm$ 39 & 0.0268 $\pm$ 0.0027 & 446 $\pm$ 37 & 0.0144 $\pm$ 0.0013 & 448 $\pm$ 39 & 0.0086 $\pm$ 0.0008\\
& & & & 546 $\pm$ 37 & 0.0165 $\pm$ 0.0016 & 546 $\pm$ 38 & 0.0083 $\pm$ 0.0009\\
& & & & & & 646 $\pm$ 39 & 0.0117 $\pm$ 0.0011\\

\hline 
\end{tabular}
}
\end{center}
{\footnotesize
{\bf Table 1.} 
The angular power spectrum $\sigma_\l^2\equiv[\l(\l+1)C_\l /2\pi]$ 
measured for the four magnitude bins.
These measurements are uncorrelated in the approximation
of Gaussian fluctuations.
Although the power spectrum is by definition non-negative, 
the allowed ranges above can include slightly negative values since
our estimators are the difference of two powers 
(total observed power minus expected shot noise power).
}
\label{DeltaTab}
\end{table*}

\subsection{Fits and priors}

As mentioned above, we need to use a prior power spectrum consistent with
the data to compute accurate error bars. To avoid the prior acquiring 
spurious wiggles caused by over-fitting noise fluctuations, 
it is desirable to use a smooth curve with as few tunable parameters
as possible that nonetheless is consistent with the final measurements.
As seen in \fig{ClFig} and \fig{ClFig2},
the simple ``concordance'' model from Wang {\etal} (2001) provides a 
good fit to the data in all four magnitude bins if we use 
bias factors $b=1.0$, $0.9$, $0.85$ and $0.8$, respectively,
so we use these power spectra as priors.
This is a flat neutrino-free model with purely scalar adiabatic fluctuations,
a cosmological constant $\Omega_\Lambda=0.66$, 
baryon density $h^2\Omega_b=0.02$, Hubble parameter
$h=0.64$ and spectral index $n_s=0.93$, normalized so that
linear $\sigma_8=0.9$ for the dark matter.
This model is well fit by a simple
untilted BBKS power spectrum (Bardeen {\etal} 1986),
parameterized by horizontal and vertical scaling factors
$\Gamma$ and $\sigma_8$ as in Szalay {\etal} (2001),
using $(\Gamma,\sigma_8)=(0.15,0.9)$.

We have corrected for non-linear evolution using the 
Hamilton {\etal} (1991) approximation as implemented in 
Jain {\etal} (1996). \Fig{ClFig} shows
nonlinear evolution to be quite 
important, especially for the brighter galaxies, with the 
corresponding linear model substantially under-predicting
the power on small scales.
In \sec{MonteSec} below, we will see that the central limit theorem nonetheless
produces a fairly Gaussian 2-dimensional projected galaxy distribution 
because of averaging along the line of sight. 

We use this cosmological model merely for a convenient parametrization of our
prior --- physical interpretation must take into account
selection function uncertainties, etc., and the reader is referred to
Dodelson {\etal} (2001) and Szalay {\etal} for a detailed treatment 
of this. The slight differences in normalization may reflect 
clustering evolution, differences in bias properties between the four samples
or some combination thereof.

On angular scales much smaller than a radian (the small-angle approximation), the
slope $n$ of a power-law angular power spectrum $C_\l$ 
is related to the power law slope $m$ of the angular 
correlation function $w(\theta)=\theta^m$ by
$m+n=-2$, so the typical power law slopes of order 
$n\sim -1.5$ in \fig{ClFig2} correspond to  
correlation function slopes of order $m\sim -0.5$, 
in good agreement with the $w(\theta)$ measurement in 
Connolly (2001).

\subsection{Relation to 3D power spectrum}

Let us conclude this section by briefly commenting on how to 
interpret our measurements.
In a companion paper (Dodelson {\etal} 2001), the present results
and those on the angular correlation function from
S2001 are used to recover an estimated 3D power spectrum 
$P(k)$. Here we present the relevant window functions that are
used as a starting point for such analyses.

\begin{figure}[tb] 
\centerline{\epsfxsize=\figsize\epsffile{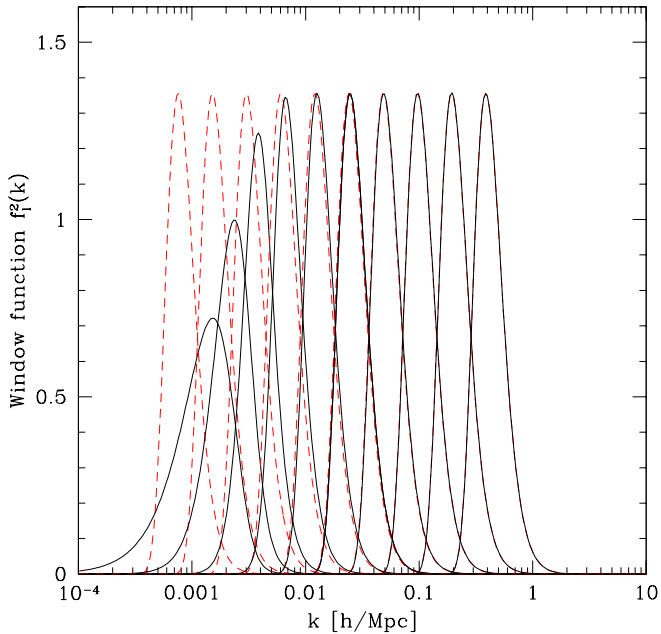}}
\caption{\label{kernelFig}\footnotesize%
The solid curves show the
exact $k$-space window function corresponding to 
$C_\l$ for multipoles
$\l=1$, 2, 4, 8, 16, 32, 64, 128, 256 and 512, respectively,
renormalized
to have unit area,
using the radial selection function for 
magnitude bin $21<r<22$.
The dashed curves show the same window functions 
computed in the small-angle approximation. These 
window functions connect the angular power
spectrum $C_\l$ to the underlying 3D power spectrum $P(k)$
via \protect\eq{3DkernelEq}.
}
\end{figure}

As described by Huterer {\etal} (2001) and Appendix B, 
the angular power spectrum
$C_\l$ is related to the 3D power spectrum $P(k)$ via the
simple relation 
\beq{3DkernelEq}
C_\l = {2\over\pi}\int_0^\infty f_\l(k)^2 P(k)k^2dk,
\eeq
where the dimensionless function 
\beq{nlEq}
f_\l(k)\equiv \int_0^\infty j_\l(kr) f(r) dr.
\eeq
Here $f$ is the probability distribution 
for the comoving distance $r$ to a random galaxy
in the survey, optionally weighted by an evolution factor, and
$j_\l$ is a spherical Bessel function.
In other words, the integral kernel transforming from 3D to our 
angular 2D case is simply a Bessel-transform of the radial 
selection function. 
A sample of these integral kernels are plotted in \fig{kernelFig}.
Accurate approximations of this kernel are available in the
small-angle limit, but we use the full expression here since it is so simple
(computational details are given in Appendix B),
and since scales where sky-curvature is non-negligible will eventually
be well probed by the SDSS.

\begin{figure}[tb] 
\centerline{\epsfxsize=\figsize\epsffile{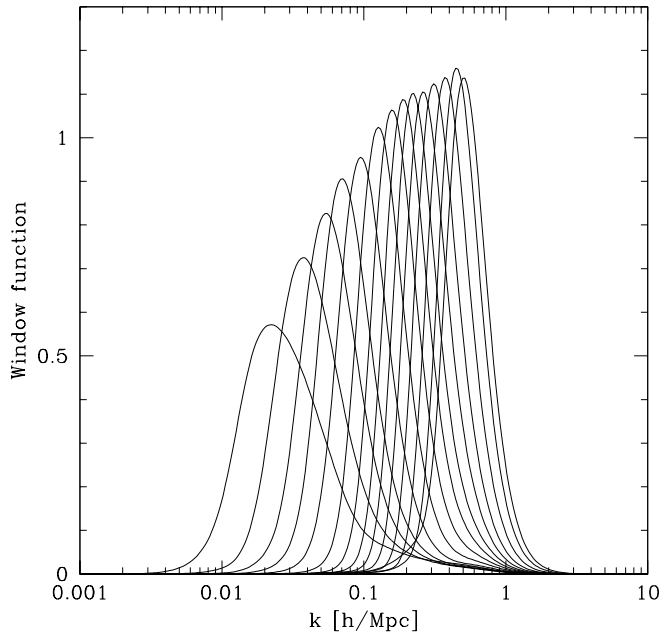}}
\caption{\label{kwindowFig}\footnotesize%
The curves show the $k$-values probed by our 14 band power measurements
for the faintest magnitude bin, thereby connecting what we measure to 
the 3D power spectrum $P(k)$.
In other words, these window functions, defined by 
\protect\eq{kwinDefEq},
are analogous to those in 
\protect\fig{lwindowFig}, but in $k$-space rather than $\l$-space.
}
\end{figure}

\begin{figure}[tb] 
\centerline{\epsfxsize=\figsize\epsffile{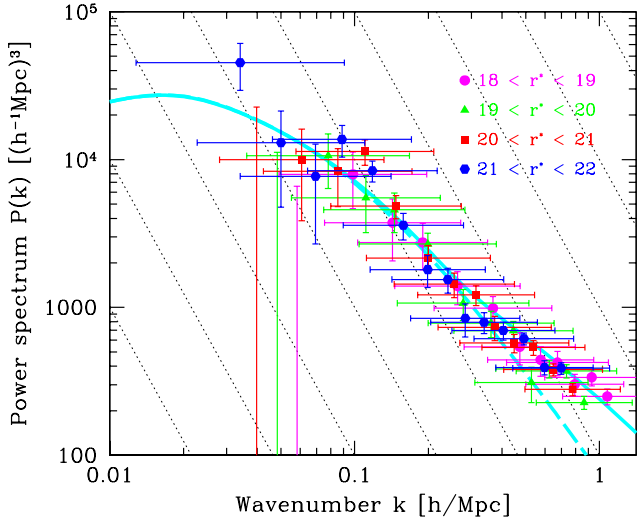}}
\vskip-0.5cm
\caption{\label{Pfig}\footnotesize%
The same band power measurements as in 
\fig{ClFig}, but plotted in $k$-space using the
window functions from 
\fig{kwindowFig}. Specifically, the
data points are the rescaled band power 
coefficients which probe a weighted average
of $P(k)$ as specified by \protect\eq{kwinDefEq}.
An inversion from 2D to 3D power spectra
is performed in Dodelson {\protect\etal} (2001)
--- the data plotted here are merely the input to
those calculations.
For comparison, the solid curve is the $\Lambda$CDM 
``concordance'' model from Wang {\protect\etal} (2001) 
with (solid) and without (dashed) nonlinear evolution.
If the mean depth has been underestimated for one of the 
Galaxy samples, the corresponding points should shift up to the left 
along the dotted lines of slope $-3$.
}
\end{figure}

By taking linear combinations of the kernels from \fig{kernelFig} 
corresponding to our $\l$-space window functions, we obtain the 
kernels of \fig{kwindowFig}, showing which $k$-values
each of our band-power measurements is probing. 
This enables us to interpret our band-powers as measuring 
weighted averages of the 3D power spectrum $P(k)$ as shown in
\fig{Pfig}. This plot is by no means a substitute for a thorough 
reconstruction of the 3D power spectrum as in Dodelson {\etal} (2001),
incorporating selection function uncertainties etc, 
but provides a useful rough guide as to which 
spatial scales are probed and, in particular, as to the $\l$-values
for each magnitude bin beyond which nonlinear clustering is likely to
be important. 

To gain further intuition about the relation between $C_\l$ and $P(k)$,
an additional approximation is instructive.
As shown in Appendix B, the 2D and 3D power spectra are approximately
related by
\beq{CPeq}
C_\l \approx {\aa\over r_*^3} P(k),
\eeq
where $k=\bb\l/r_*$ and $r_*$ is the mean spatial depth of the survey.
The key approximation made here is that $C_\l$ in fact 
probes not simply the power $P$ at wavenumber $k$, 
but rather a weighted average of $P$  
with a window function of width $\Delta k\approx\cc k$.
Here $\aa$, $\bb$ and $\cc$ are dimensionless constants of order unity
that depend only on the shape of the radial selection function, not on its
depth. For the SDSS case described in Dodelson {\etal} (2001), 
$\aa\approx 0.75$, $\bb\approx 1.1$ and the smoothing width 
$\cc=\Delta k/k\approx 0.37$.

In other words, we can interpret $C_\l$ as a smoothed version
of $P(k)$ shifted vertically and horizontally in a log-log plot such 
as \fig{ClFig2}. 
Moreover, \eq{CPeq} shows that mis-estimates of the 
radial selection function depth $r_*$ 
will simply shift the entire $P(k)$-curve 
along the lines of slope $-3$ shown in \fig{Pfig},
without changing its shape.

\section{Robustness and limitations of results}
\label{SystematicsSec}

How reliable are the angular power spectrum measurements
computed above?
In this section, we discuss the 
underlying assumptions and their limitations.
We focus on three areas and discuss them in turn:
potential problems with the input data, potential problems with the 
data processing (analysis algorithms/software) and potential 
problems with underlying assumptions, notably Gaussianity.

\subsection{Issues related to the input data}

The input data used in our analysis have been extensively tested for
potential systematic errors by S2001, and constitute arguably
the cleanest deep angular survey data to date. In particular,
S2001 present a battery of tests for problems involving 
star-galaxy separation and modulation of the galaxy detection
efficiency by external effects such as 
photometric calibration, seeing conditions and 
Galactic extinction. By cross-correlating the galaxy
maps with various two-dimensional ``trouble templates'' corresponding to
variations in seeing, reddening, stellar density, camera column
structure, \etc, the various effects were quantified and reduced
to negligible levels by sharpening the seeing and reddening cuts.
This gives us confidence that the errors in our star-galaxy separation
algorithm (which depends on seeing) and reddening estimates have
negligible effect on our estimates of angular power spectra even in the 
faintest magnitude bin.

As an additional precaution, we complement the tests from S2001
with three that are tailored for our $C_\l$-analysis.
Specifically, we compute the angular power spectra of the seeing and
reddening templates, which were found to be
the most serious challenges in S2001, and with a 
photometric calibration error template.
Strictly speaking, these of course do not {\it have}
well-defined power spectra, since they are not
isotropic random fields. Rather, what is relevant here is the amplitude and
shape of the bias that they would add to our estimates of
the galaxy power spectra. We therefore process these templates in exactly
the same way as the galaxy maps, with the pair-weightings (the $\Q_i$-matrices)
given by the {\it galaxy} noise and signal matrices.
We use the weighting and sky mask corresponding to the faintest
magnitude bin, since this is the one that is most vulnerable to 
these systematics --- both because these galaxies have the poorest 
signal-to-noise ratio in the CCD photometry 
(Lupton {\etal} 2001) and because they have 
the lowest intrinsic angular clustering amplitude.

We use the same seeing and reddening templates as S2001, \ie,
the second moment of the point-spread 
function for each pixel and the extinction
correction from Schlegel {\etal} (1998).
In order to provide a meaningful comparison between the amplitudes
of signal and systematics, we need to estimate the conversion factor
from seeing or reddening power to galaxy fluctuation power.
We do this using the cross-correlations presented in figures
8 and 9 of S2001. To be conservative and err on the side of caution,
we use the relevant $2\sigma$ cross-correlation 
upper limits, 0.0017 and 0.0038,
for seeing and reddening, respectively. 
These values are the largest upper limit on any angular scale,
but we have used them at all angular scales $\l$ to be conservative.

\begin{figure}[tb] 
\centerline{\epsfxsize=\figsize\epsffile{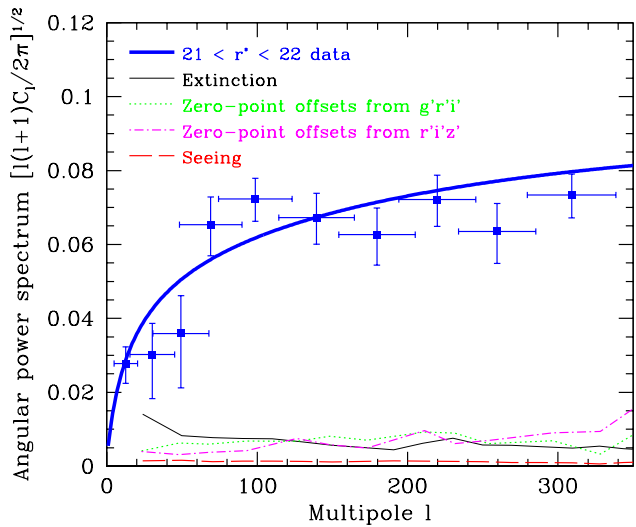}}
\vskip-0.5cm
\caption{\label{systematicsFig}\footnotesize%
To give a feeling for the magnitude of potential systematic
errors, the power spectrum in the faintest 
(and most vulnerable) magnitude bin is compared
with the power spectra for seeing, extinction and photometry related
modulations of the observed galaxy density
under very pessimistic assumptions.
}
\end{figure}

The corresponding angular power spectra for seeing and extinction are shown in 
\fig{systematicsFig} and, as opposed to the galaxy fluctuations,
they are seen be  flat or rise towards {\it larger} angular scales.
For the reddening case, this is in good agreement with the findings 
of Vogeley (1998) and measurements of the dust 
power spectrum. The combined DIRBE and IRAS dust maps
suggest a power law $C_\l\propto\l^{-2.5}$
(Schlegel {\etal} 1998), and a recent analysis of the 
DIRBE maps has supported an even redder slope
with an $\l^{-3}$ power law for $\l\simlt 300$ (Wright 1998).

As a template for photometric calibration errors, we identify a
feature of the stellar distribution in color space and measure it as a
function of position in the sky.  As seen, e.g., in Finlator {\etal}
(2000), the locus of stars in the $g-r,r-i$ color plane shows two
branches: stars cooler than $\sim$M0 have almost constant $g-r$
colors, while hotter stars show a strong correlation between the $g-r$
and $r-i$ color.  The crossing point of linear fits to the stellar
locus in these two branches should be independent of position on the
sky, thus variations in this crossing point are a sensitive measure of
photometric calibration errors in $g, r$ and $i$.  Similarly, the
stellar locus in the $r-i,i-z$ plane is almost linear; one can define
the $i-z$ color corresponding to the $r-i$ color of the crossing point
measured from the $g-r,r-i$ plane.  

We have measured the crossing colors from the stars in our sample on
scales of two degrees by 13 arcminute (the width of a scanline), and
attribute all observed variations to errors in $r^*$ to be
conservative; the distribution of the implied error is roughly
Gaussian, with a sigma of 0.015 magnitudes. We convert these
$r^*$-fluctuations into density fluctuations by multiplying by the
source count slope $d\ln N/dr$. This slope is of order unity at
$r^*=19$ and flattens at fainter magnitudes (Yasuda {\etal} 2001), so
we make the conservative assumption $d\ln N/dr=1$.  The angular power
spectra of these two calibration error maps are shown in
\fig{systematicsFig} and are seen to be approximately flat (scale-invariant).

the almost horizontal part towards the left, and stars later than 
$\sim$M0 are found in the vertical branch with $g'-r'\sim 1.4$

It is reassuring that even with the extremely pessimistic assumptions
described above, the expected contaminant signals remain much smaller than
the observed galaxy power spectrum
all the way
out to the largest scales currently probed.
Extrapolation to extremely large scales suggests
that even extinction should remain subdominant for  $\l\simgt 5-10$.

In summary, we have found that systematics are small 
even in the nearly worst-case scenario shown in \fig{systematicsFig}.
Moreover, the SDSS will provide other internal checks on many systematics, notably
the extinction correction, so these are unlikely to prove
a significant limitation on determining the power spectrum.

\subsection{Issues related to algorithms, software and assumptions}
\label{MonteSec}

Since our analysis consists of a number of somewhat complicated steps,
it is important to test the integrity of both the software and the underlying methods.
We do this using two types of Monte Carlo simulations:
\begin{enumerate}
\item We analyze 1000 Monte-Carlo maps $\x$ that are 
drawn from a multivariate Gaussian distribution with 
vanishing mean and covariance matrix $\C$.
\item We analyze 100 Monte-Carlo galaxy samples including non-linear
clustering as described in Scoccimarro \& Sheth (2001) and S2001.
\end{enumerate}
Both sets of mock data were processed through our
analysis pipeline, enabling us to check not only whether 
we obtained the correct answer
on average, but also whether the scatter and the error correlations
corresponded to the predicted values.
The first suite of Monte Carlos offered precision end-to-end tests
of the algorithms and the software, since 
errors or bugs in any of the
many intermediate steps would have manifested themselves here. They used the exact same 
survey geometry as the real data, including the seeing and reddening
masks of S2001.

\begin{figure}[tb] 
\centerline{\epsfxsize=\figsize\epsffile{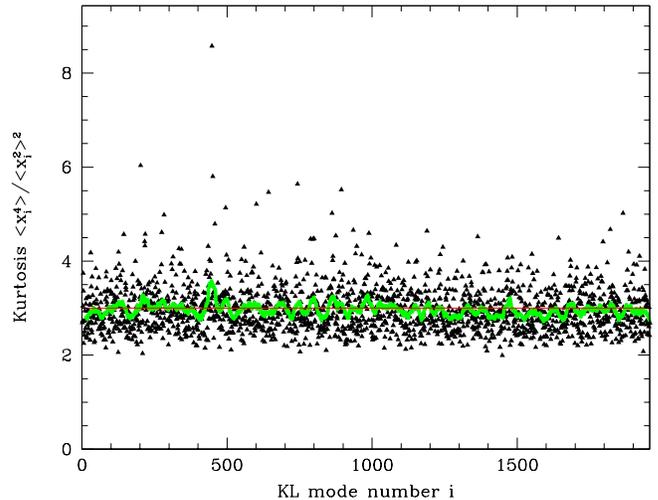}}
\vskip-0.5cm
\caption{\label{kurtosisFig}\footnotesize%
The simulated kurtosis of the first 1957 KL-coefficients is shown for
the $18<r<19$ magnitude bin. Each triangle represents the kurtosis 
of the distribution for the corresponding mode as 
measured from the 100 nonlinear Monte Carlo simulations
described in the text. The thick curve shows a running
average of 25 modes, whereas the thin horizontal line 
shows the Gaussian prediction of three.
}
\end{figure}

The second suite of Monte Carlos provides a way of quantifying 
the limits of applicability of 
the Gaussian assumption.
They were constructed using the PTHalos code (Scoccimarro \& Sheth 2001)
as described in detail in Scranton et al (2001), covering
a rectangular $90^\circ\times 2.5^\circ$ sky region. In short, this code is
a fast approximate method to build non-Gaussian density fields with
realistic correlation functions, including non-trivial galaxy biasing
(obtained by placing galaxies within dark matter halos with a prescribed
halo occupation number as a function of halo mass).

The non-Gaussian effects produced by non-linear evolution
encode information that can be captured by measuring 
higher-order moments and other statistics.
Since this route is explored in detail in Szapudi {\etal} (2001), 
we will not pursue it here. 
However, we need to quantify the level to which this non-Gaussianity
affects our results.

Since our power spectrum estimates are simply quadratic functions of
the density field, they give unbiased measurements of the 
underlying power spectrum even if 
the fluctuations are non-Gaussian.
In other words, our calculation of window functions,
KL-modes {\etc} is completely
general and does not make any assumptions about Gaussianity.
The one place in the quadratic estimator formalism where Gaussianity is
assumed is in the computation of error bars. Since
the variances of our power spectrum estimates involve
fourth moments of the observed density fields (kurtosis), 
they will generally
differ from the Gaussian prediction in the presence of
non-Gaussianity --- typically by being larger. 
The covariance between band power estimates likewise involves
fourth moments, so we should not expect our error bars to retain
their attractive property of being uncorrelated down into the
nonlinear regime. The third moment (skewness) of the galaxy distribution
also affects the power spectrum error bars via coupling to the Poissonian
shot noise, at a level of the same order of magnitude as the kurtosis.

To quantify this effect, \fig{kurtosisFig}
shows the kurtosis of the first 
1957 KL-coefficients for the $18<r<19$ magnitude bin. 
(This brightest magnitude bin is expected to be
the most non-Gaussian, both because it probes the smallest 
spatial scales and because it involves the least amount
of line-of-sight averaging --- such averaging makes the density field
mory Gaussian as per the central limit theorem .)
The kurtosis was computed by processing 
the 100 nonlinear Monte Carlo simulations
through our analysis pipeline and computing the
variance and fourth moment of the 100
values obtained for each mode.
The dimensionless kurtosis plotted is the
fourth moment divided by the square of the variance,
\ie, $\expec{y_i^4}/\expec{y_i^2}^2$, and would equal three
for Gaussian fluctuations (in which case the KL-coefficients
would be simply independent Gaussian random variables).
Since we have only 100 simulations, there is still a
fair amount of scatter. To further reduce the scatter, 
we have therefore added a line in \fig{kurtosisFig}
showing a running average of 25 consecutive triangles.
The scatter (which is determined by eighth 
moments) appears to rise somewhat initially, as modes
probe progressively smaller angular scales, then 
decreases again as cosmological fluctuations become smaller than
Poisson shot noise fluctuations.
However, the kurtosis itself, which (with the skewness) is the 
only quantity that
affects our error bars, is seen not to depart significantly
from the Gaussian value on any of the angular scales we have probed.
This implies that non-Gaussianity does not appear to
have a major impact on our results. This is partly by design, since
we chose to focus our analysis on the largest scales.

In other words, although non-Gaussian effects are very strong on small
scales (indeed, the onset of non-linear evolution is evident in \fig{ClFig}), 
they have only a weak effect on the error bars of 
our large scale angular power spectrum. 
Factors contributing to this are the dominance of shot
noise on small scales (on a mode-by-mode basis) as well as the central limit theorem,
suppressing non-Gaussianity by averaging fluctuations along the
line of sight.

In conclusion, we have computed the large-scale angular power spectrum
from early SDSS data and performed a series of tests validating our results.
The cosmological implications of our measurements are discussed in a 
companion paper by Dodelson {\etal} (2001).
Although these results are interesting in their own right, 
perhaps the most important conclusion is that the 
lack of discernible systematic errors 
even on scales as large as tens of degrees bodes extremely well
for analysis of future SDSS data.
The present data covered about 160 square degrees, \ie, less
than $2\%$ of the full survey that will eventually be available,
so angular clustering studies are likely to remain at the
forefront of the quest for a detailed understanding of 
cosmic clustering.

\bigskip

The authors wish to thank Andrew Hamilton and Lloyd Knox for helpful comments.
%

The Sloan Digital Sky Survey (SDSS) is a joint project of The University of
Chicago, Fermilab, the Institute for Advanced Study, the Japan Participation
Group, The Johns Hopkins University, the Max-Planck-Institute for Astronomy
(MPIA), the Max-Planck-Institute for Astrophysics (MPA), New Mexico State
University, Princeton University, the United States Naval Observatory, and the
University of Washington. Apache Point Observatory, site of the SDSS telescopes,
is operated by the Astrophysical Research Consortium (ARC).

Funding for the project has been provided by the Alfred P. Sloan Foundation, the
SDSS member institutions, the National Aeronautics and Space Administration, the
National Science Foundation, the U.S. Department of Energy, the Japanese
Monbukagakusho, and the Max Planck Society. The SDSS Web site is
http://www.sdss.org/.
MT was supported by NSF grant AST00-71213,
NASA grants NAG5-9194 and NAG5-11099,
the University of Pennsylvania Research Foundation
and the David and Lucile Packard Foundation.
SD is supported by the DOE and
by NASA grant NAG5-7092 at Fermilab, and by NSF Grant PHY-0079251.
DJE was supported by NASA through Hubble
Fellowship grant \#HF-01118.01-99A from the Space Telescope Science
Institute, operated by the AURA Inc., under NASA contract NAS5-26555.

\appendix

\def\lowercasek{k}
\def\lowercasew{w}
\section{The relation between}
\vskip-0.3cm
\centerline{DIFFERENT QUADRATIC ESTIMATORS}
\bigskip

The purpose of this Appendix is to describe the $\Q$-matrices
that define our analysis as well as to elucidate the relationship 
between quadratic estimators of $C_\l$, $w(\theta)$ and $P(k)$.
From an information-theoretic point of view, we will see
that the key issue is not which of the three functions
one tries to measure, but what pair weighting is used in the process
--- the minimum-variance weighting retains all information about all three
of them in the Gaussian approximation.
Indeed, we will see that the decorrelated minimum-variance estimators of all
three functions are one and the same set of numbers, just
normalized differently!

\subsection{The $\Q$-matrices used in our analysis}

As described in \sec{ResultsSec}, our power spectrum estimators
are quadratic functions of the observed galaxy density.
The estimator of the power in the $\ith$ band is therefore defined 
by a symmetric matrix $\Q_i$ that gives the weight assigned to each pair
of pixels (or KL-coefficients) via \eq{qDefEq}.
In our analysis, we make the choice
\beq{QdefEq}
\Q_i = \sum_j (\B)_{ij}\C^{-1}\P_j\C^{-1}
\eeq
for a $50\times 50$ matrix $\B$ that will be defined below.
It can be shown (Tegmark 1997) that this choice 
distills all the cosmological information from the original galaxy map
$\x$ into the 
(much shorter) vector $\phat$
in the approximation of Gaussian fluctuations, 
as long as the matrix $\B$ is invertible and 
the binning scale $\Delta\l$ is narrower than
the scale on which the power spectrum varies substantially.
In this approximation, 
the mean and covariance of the 
quadratic estimator vector $\phat$ defined by \eq{qDefEq}
is given by 
\beqa{simpleQmomentseq}
\expec{\phat}&=&\B\F\p,\\ \label{MdefEq}
\M\equiv\expec{\phat\phat^t}-\expec{\phat}\expec{\phat}^t&=&\B\F\B^t,
\eeqa
where
\beq{GaussFisherEq}
\F_{ij} = {1\over 2}\tr
\left[\C^{-1}\P_i\C^{-1}\P_j\right]
\eeq
is the so-called Fisher information matrix.
As advocated in Tegmark \& Hamilton (1998), we choose
$\B=\D\F^{-1/2}$, where $\D$ is a diagonal matrix 
whose elements are chosen so that 
the window matrix $\W=\B\F$ has unit row sums.
This choice has the virtue of giving uncorrelated error bars
(the covariance matrix of \eq{MdefEq} becomes the
diagonal matrix $\M=\D^2$) and
narrow, well-behaved window functions as seen in \fig{lwindowFig}.

\subsection{The relation between quadratic estimators of
$C_\l$, $\lowercasew(\theta)$ and $P(k)$}

Suppose the angular power spectrum parameters $p_i$ can be expressed
as linear combinations of some other parameters $p'_i$, \ie,
\beq{ppDefEq}
\p = \A\p'
\eeq
for some matrix $\A$. There are two such interesting examples, involving
$w(\theta)$ and $P(k)$, respectively.
If we define $p'_i\equiv w(\theta_i)$, \ie, the angular correlation function
amplitude in the $\ith$ angular bin, then $\A$ is given by 
\beq{AEq1}
\A_{\l i} = {1\over 2\pi} P_\l(\cos\theta_i)\sin\theta_i \Delta\theta,
\eeq
where $P_\l$ is a Legendre polynomial and 
$\Delta\theta$ is the width of the angular bins.
If we define $p'_i\equiv P(k_i)$, \ie, the 3D power spectrum
in the $\ith$ $k$-bin, then $\A$ is given by 
\beq{AEq2}
\A_{\l i} = K_\l(k_i) k_i^3 \Delta\ln k,
\eeq
where $K_\l$ is given by \eq{ExactKernelEq}
and 
$\Delta\ln k$ is the width of the (logarithmic) $k$-bins.
(Throughout this subsection, we assume for simplicity the $\theta$- or $k$-bins 
are narrow enough to resolve any features in $w(\theta)$ or $P(k)$, and that 
there is no $\l$-binning, defining $p_\l=C_\l$.)

Using \eq{ppDefEq},
we can construct quadratic estimators $\phat'$ to measure 
$\p'$ directly, without going through the intermediate step of
measuring the angular power spectrum $\p$ first. 
Writing $\S=\sum p'_i P'_i$ by analogy with \eq{CdefEq},
the new $\P$-matrices are given in terms of the old ones by
\beq{PprimeDefEq}
\P'_i = \sum_j \A^t_{ij}\P_j.
\eeq
Using equations\eqn{qDefEq} and\eqn{QdefEq} 
therefore shows that the new estimators are related to the old ones by
\beq{oldnewRelationEq1}
\phat' = \B'\A^t\B^{-1}\phat.
\eeq
Here $\B'$ is the $\B$-matrix corresponding to the
new parameters $\p'$, and we use the same notation with primes $'$
for other matrices below.
To obtain an intuitive understanding for this relation, let 
us simplify things by using the choice $\B\equiv\D\L^{-1}$ in place of our
previous choice $\B\equiv\D\F^{-1/2}$, where $\L$ is the lower-triangular
matrix obtained by Cholesky-decomposing the Fisher matrix as $\F=\L\L^t$.
$\L$ can be viewed of as simply an alternate choice of square root of $\F$.
As described in Tegmark \& Hamilton (1998), this choice has the same desirable properties
as $\B\equiv\D\F^{-1/2}$ except that it gives asymmetric window functions
($\F^{1/2}$ is symmetric whereas $\L$ is not).
A straightforward calculation shows that $\F'=\A^t\F\A$, so 
$\L'=\A^t\L$ and \eq{oldnewRelationEq1} reduces to 
\beqa{oldnewRelationEq2}
\phat' 
&=& (\D'{\L'}^{-1})\A^t(\D\L^{-1})^{-1}\p\nonumber\\
&=& (\D'\L^{-1}\A^t)\A^t(\L\D^{-1})\p
= \D'\D^{-1}\p,
\eeqa
a diagonal matrix.
In other words, if we use the same number of $\l$-values as there are 
bins (for $\theta$ or $k$), with $\A$ an invertible square matrix,
then the old estimators $\ph_i$ and the new estimators $\ph'_i$ are
the exact same numbers except differently normalized!
The normalization factors $\D_{ii}$ and $\D'_{ii}$
simply let us interpret the measurements as probing weighted
averages of $\p$ and $\p'$, respectively.

This shows that there is no fundamental difference between measuring 
$C_\l$, $w(\theta)$, $P(k)$ or some other linear transformation of the 
power spectrum with quadratic estimators of the form of 
\eq{QdefEq}. Not only do they all contain the same information
(keeping the $\P$-matrices the same, 
two different $\phat$ computed with different $\B$-matrices are 
trivially related by $\phat'=\B'\B^{-1}\phat$),
but even the $\B$-matrices will be essentially the same if
we decorrelate the measurements. This means that the
rescaled $C_\l$-estimates shown in \fig{Pfig} can alternatively be 
interpreted as decorrelated quadratic estimators of $P(k)$,
or as rescaled decorrelated quadratic estimators of $w(\theta)$! 
The reason that this paper purports to measure
$C_\l$ rather than $w(\theta)$ is simply that 
the window-functions for our estimators
turn out to be narrow and well-behaved in $\l$-space,
but wide and partially negative in $\theta$-space.

\subsection{The relation between different pair weightings}

In the companion paper by Connolly {\etal} (2001), the angular 
correlation function $w(\theta)$ was measured with a different technique,
using so-called Landy-Szalay (LS) estimators.
LS-estimators are also quadratic estimators, and in our notation corresponds 
to replacing the $\Q$-matrix choice of \eq{QdefEq} by 
\beq{LSQdefEq}
\Q_i = N_i\P_i.
\eeq
For the $w(\theta)$-case, the $\P$-matrices take the simple form
$\P_{jk}=1$ if the angular separation between pixels $j$ and $k$ falls 
in the angular bin around $\theta_k$, vanishing otherwise.
The normalization constants $N_i$ are simply the number of pixel pairs 
with angular separation in the $\ith$ angular bin, so $N_i=\tr\P_i^2$.
The estimators corresponding to \eq{LSQdefEq} are not simply related 
to those corresponding to \eq{QdefEq} since the $\C^{-1}$-weighting is 
absent. They therefore do not contain the same cosmological information.
However, they have two other very desirable properties.
The first is that the window matrix from \eq{qMeanEq} is
\beq{LSwindowEq}
\W_{ij}\equiv\tr[\P_i\Q_j]
=N_i\tr[\P_i\P_j]
=\delta_{ij}, 
\eeq
the identity matrix.
This means that the LS quadratic estimators can be interpreted as 
exact measurements of $w(\theta)$ with no smoothing whatsoever.
The second advantage is computational speed. Since no time-consuming 
matrix inversions are necessary, the LS-estimator of $w(\theta)$ can be 
computed with many more pixels than would otherwise be feasible, 
probing the clustering down to far smaller angular scales than we have 
probed in this paper.
Finally, it is worth noting that the lossless property of the
quadratic estimators of \eq{QdefEq} breaks down on small scales where
fluctuations become non-Gaussian, making the computationally superior 
LS-estimators preferable in this regime.
The bottom line is that the $C_\l$-estimation used here 
and the LS-estimation of $w(\theta)$ used in 
Connolly {\etal} (2001) are highly complementary approaches,
being preferable on large and small angular scales, respectively.

\section{The relation between $C_\l$ and $P(\lowercasek)$}

In this Appendix, we discuss the close relation between 
the angular power spectrum $C_\l$ that we have measured and
the underlying 3D power spectrum $P(k)$.
The purpose is both to review their exact quantitative 
relation, and to provide qualitative intuition for 
this relation and how it is affected by mis-estimates 
of the radial selection function.
As we will see, $C_\l$ can be interpreted as essentially a smoothed
version of $P(k)$ shifted horizontally and vertically on a log-log plot,
and a mis-estimate of the mean survey depth would shift the power spectrum
along lines of slope $-3$.

\subsection{The exact relation}

The angular power spectrum $C_\l$ is related to the 
3D power spectrum $P(k)$ by 
\beq{KernelDefEq}
C_\l = \int_0^\infty K_\l(k) P(k)k^2dk
\eeq
for a dimensionless integral kernel $K_\l(k)$ that depends on the radial selection function of
the survey.
As shown in Appendix A of Huterer {\etal} (2001), 
\beq{ExactKernelEq}
K_\l(k) = {2\over\pi} f_\l(k)^2,
\eeq
where $f_\l$ is the Bessel transform of 
the radial selection function $f(r)$ as given by \eq{nlEq}.
Specifically, $f(r)=g(r)h(r)$, where $g(r)$ is the probability distribution for the
{\it comoving} distance from us to a random galaxy in the survey
and $h(r)$ is an optional (bias and clustering) evolution term of order unity, so 
$g$ has units of inverse length and $h$ is dimensionless.
Defining $\Nbar$ to be the expected number of galaxies within a sphere of a certain
radius, we thus have 
\beq{fEq}
g(r)\propto {d\Nbar\over dr} 
= {d\Nbar/dz\over dr/dz} 
= {H(z)\over H_0 r_0}{d\Nbar\over dz},
\eeq
normalized so that 
\beq{NormEq}
\int_0^\infty g(r)dr=1.
\eeq
Here 
$r_0\equiv c/H_0 \approx 3000 h^{-1}$Mpc, 
and the relative Hubble parameter is
\beq{Heq} 
{H(z)\over H_0} = \sqrt{\Ol + (1-\Ol-\Om)(1+z)^2 + \Om(1+z)^3}
\eeq
for a cosmology with density parameters
$\Om$ and $\Ol$ for matter and vacuum energy, respectively.
This means that uncertainties about $d\Nbar/dz$ and uncertainties about
the cosmological parameters $(\Om,\Ol)$ get combined, entering only via 
the single function $g(r)$.

The evolution term $h(r)$ relates the past and present galaxy clustering amplitudes,
and is given by 
\beq{hEq}
h = [P(k;z)/P(k)]^{1/2}
\eeq
for a flat Universe.
If space should turn out to be curved despite present evidence to the contrary,
$h$ gets multiplied by a correction factor as in Peebles (1980).
The factor $h$ is likely to remain close to unity for the low redshifts 
$z\simlt 0.5$ probed by the SDSS, especially since the effects of bias 
evolution and dark matter clustering evolution 
appear to partially cancel (Blanton {\etal} 2000).
The clustering evolution is expected to be small over this redshift range since 
linear growth grinds to a halt at recent times when vacuum energy becomes dominant. 
Rather than attempting a complicated and poorly justified
model for $h(r)$, we therefore simply set $h(r)=1$ and reinterpret the measured
$P(k)$ as the power spectrum at the effective redshift corresponding
to $r\sim \l/k$ (Dodelson {\etal} 2001).

In practice, we evaluate the Bessel-transform of \eq{nlEq} for 
512 logarithmically equispaced $k$- and $r$-values
using Fourier methods from the FFTlog package of Hamilton (2000).
This is an efficient $N\,$log$\,N$ algorithm, evaluating all kernels 
up to $\l=1000$ in about a minute on a workstation.
Sample results are shown in \fig{kernelFig}
using the selection function for the $21<r<22$ band described
by Dodelson {\etal} (2001) for a Universe with
$\Om=0.3$, $\Ol=0.7$.

\subsection{The small-angle approximation}

The approximation 
\beq{ApproxKernelEq}
K_\l(k) \approx {1\over\l k^2} f\left({\l\over k}\right)^2,
\eeq
becomes accurate in the small-angle limit
(see, \eg, Kaiser 1992; Baugh \& Efstathiou 1994),
as illustrated in \fig{kernelFig}. This is the
$\l$-space version of Limber's equation, which relates
$P(k)$ to the angular correlation function $w(\theta)$.
\Eq{ApproxKernelEq} can be derived directly from
\eq{ExactKernelEq} by noting that for large $\ell$, 
the spherical Bessel function $j_\l(kr)$ becomes sharply 
peaked around $kr=\l$. Assuming that $f(r)$ is a smoothly 
varying function relative to this peak width, we can thus 
approximate it by $f(\l/k)$ and take it out of the integral
in \eq{nlEq}, obtaining
\beq{nlApproxEq} 
f_\l(k)\approx f\left({\l\over k}\right) \int_0^\infty j_\l(kr) dr
\approx \left({\pi\over 2 k^2\l}\right)^{1/2} f\left({\l\over k}\right),
\eeq
since 
\beq{BesselIntEq}
\int_0^{\infty} j_\l(x)dx 
= {\sqrt{\pi}\over 2}
{\Gamma\left({\l+1\over 2}\right)\over\Gamma\left({\l+2\over 2}\right)}
\approx\sqrt{\pi\over 2\l}
\eeq
for $\l\gg 1$.

\subsection{$\lowercasek$-space window functions}

Since our measured band powers probe linear combinations of the
actual power spectrum coefficients $C_\l$, and these
in turn are linear combinations of $P(k)$, we can  
reinterpret our band-power measurements $\ph_i$ as 
probing $P(k)$ directly.
In other words, the window matrix $\W$ from 
\eq{qMeanEq} relates our measurements to $C_\l$
and the kernel $K_\l(k)$ of \eq{KernelDefEq}
relates $C_\l$ to $P(k)$, so combining the two relates 
our measurements to $P(k)$. Specifically, 
these two equations give 
\beq{kWinEq1}
\expec{\ph_i} = \int_0^\infty \sum_\l W_{i\l} K_\l(k) P(k) k^2 dk
\eeq
for the case of no $\l$-binning. Since we have binned 
our angular power spectrum in $\l$-bins of width $\Delta\l=20$,
the sum over $\l$ in \eq{kWinEq1} gets replaced by a sum over 
bins and $K_\l(k)$ gets replaced by its average over each $\l$-bin.

\Eq{kWinEq1} is seen to take the simple form
$\expec{\ph_i} = \int W_i(k) P(k) d\ln k$ for functions 
$W_i(k)\equiv\sum_\l W_{i\l} K_\l(k) k^3$ that
are never negative. Defining normalization constants
$c_i\equiv\int W_i(k)d\ln k$,
this means that we can interpret
our measurements $\ph_i$ as probing simply $c_i$ times weighted averages
of $P(k)$ with weight functions $W_i(k)/c_i$.
However, caution is necessary before using this fact to make plots
like \fig{Pfig}. The reason is that if the window functions $W_i(k)/c_i$ are wide
(which they are) and the function to be measured varies substantially
on the scale of this window (which $P(k)$ typically does), then
the weighted average will be dominated by one edge of the window.
For instance, in the regime of \fig{Pfig} where $P(k)$ is rapidly falling,
the integral $\int W_i(k) P(k) d\ln k$ would be dominated by the contribution
from $k$-values leftward of the peak of $W_i(k)$, causing the corresponding point
in \fig{Pfig} to be plotted misleadingly far to the right.
Such problems can be avoided by redefining the quantity to be measured to be a roughly 
constant function. 
This is why we chose to measure $\sigma_\l^2$ rather than $C_\l$ above. 
Following, \eg, Eisenstein \& Zaldarriaga (2000) and
Hamilton {\etal} (2000), we therefore interpret our measurements as weighted averages
of the {\it relative} power spectrum, defined as $P(k)/P_*(k)$, where $P_*(k)$ is our
fiducial power spectrum described in \sec{ResultsSec}.
This relative power will be a fairly constant function (of order unity)
as long as the shape of our fiducial power spectrum is not grossly inconsistent
with the truth.
We therefore write
\beq{kwinDefEq}
\expec{\ph_i} = \int W_i(k) {P(k)\over P_*(k)} d\ln k,
\eeq
where we have defined window functions
\beq{kWinDefEq}
W_i(k)\equiv P_*(k)k^3\sum_\l W_{i\l}K_\l(k),
\eeq
These functions are plotted in \fig{kwindowFig}
for the faintest magnitude bin.
This equation is analogous to 
\eq{qMeanEq}, linking our measurements to 
the 3D power spectrum $P(k)$ rather than the angular spectrum
$C_\l$.
The rescaled band-power coefficients $\ph_i/c_i$ are thus weighted averages
of the relative power spectrum, where 
$c_i\equiv\int W_i(k)d\ln k$ as before.
The numbers $(P_*(k)/c_i)\ph_i$ can therefore be viewed as a measurements of $P(k)$,
and are plotted in \fig{Pfig} at the $k$-values corresponding to the means of the
distributions $W_i(k)$, with horizontal bars indicating the 
rms widths of $W_i(k)$. 
The results are seen to be roughly consistent 
between magnitude bins and
in agreement with a standard $\Lambda$CDM 
power spectrum.

\subsection{The narrow window approximation and the poor man's Limber inversion}

Let us now make an approximation aimed at building qualitative intuition 
for how $C_\l$ is related to $P(k)$ and, in 
particular, for how this relation depends on the details of the selection 
function $f(r)$.
\Fig{kernelFig} shows that the kernels $K_\l(k)$ from \eq{KernelDefEq} are
fairly narrow positive functions with a single peak. 
Defining their $n^{\rm th}$ moments as
\beq{MomentDefEq}
\expec{k^n}_\l\equiv \int K_\l(k) k^n d\ln k,
\eeq
they are therefore roughly characterized by their areas 
$A_\l$, means $k_\l$ and rms widths $\Delta k_\l$ given by
\beq{kMomentEq2}
A_\l\equiv\expec{k^0}_\l, \quad
k_\l\equiv{\expec{k^1}_\l\over A_\l}, \quad {\rm and} \quad
\Delta k_\l\equiv\left({\expec{k^2}_\l\over A_i}-k_\l^2\right)^{1/2},
\eeq
respectively.
In the crude approximation that the widths $\Delta k_\l$ of these 
curves are smaller than the scale on which the dimensionless power
$k^3 P(k)$ varies appreciably, 
we can approximate $k^3 P(k)$ by $k_\l^3 P(k_\l)$ in 
the integral of \eq{KernelDefEq}, obtaining simply
\beq{CPeq2}
C_\l 
\approx \int_0^\infty K_\l(k) P(k_\l)k_\l^3 d\ln k
= A_\l k_\l^3 P(k_\l).
\eeq
Since $P(k)$ is roughly a power law near any given $k$, this 
approximation is accurate in the limit where $\Delta k_\l/k_\l\ll 1$.
 
To highlight the scale dependence of the problem,
let us define the mean comoving distance by
\beq{rstarEq}
r_*\equiv\int f(r)r dr 
\eeq
and the dimensionless probability distribution function (PDF) 
for $x\equiv r/r_*$ by
\beq{fstarEq}
f_*(x) \equiv r_* f(r_* x).
\eeq
The PDF $f$ thus has both area and mean of unity, and quantifies
only the {\it shape} of the radial selection function, with $r_*$
encapsulating the physical depth of the survey.
Since $f(r)=f_*(r/r_*)/r_*$, 
substituting the small angle approximation of \eq{ApproxKernelEq}
into \eq{MomentDefEq} now gives
\beq{kMomentEq3}
\expec{k^n} \approx {\alpha_n\l^{n-3}\over r_*^n},
\eeq
where the dimensionless constants 
\beq{AlphaEq}
\alpha_n\equiv\int f_*(x)^2 x^{1-n} dx
\eeq
are all of order unity. Substituting \eq{kMomentEq3}
into \eq{kMomentEq2} now gives the simple results
\beqa{kMomentEq4}
A_\l&=&{\alpha_0\over\l^3},\nonumber\\
k_\l&=&{\bb\l\over r_*},\\
\Delta k_\l&=&\cc k_\l,\nonumber
\eeqa
where  
$\bb\equiv\alpha_1/\alpha_0$ and
$\cc\equiv (\alpha_0\alpha_2/\alpha_1^2-1)^{1/2}$ are again
dimensionless constants of order unity that depend only on the
shape function $f_*$. Substituting \eq{kMomentEq4} into
\eq{CPeq2} thus gives the extremely simple formula of 
\eq{CPeq}, where $\alpha\equiv\alpha_0\beta^2=\alpha_1^3/\alpha_0^2$.
\Eq{CPeq} tells us that we can perform a ``poor man's Limber inversion''
by simply making a log-log plot of $C_\l$ and changing the
axis labels to $k$ and $P(k)$, respectively, making the substitutions
\beq{SubstitutionEq}
\l\mapsto k={\beta\l\over r_*},\quad C\mapsto P = {C\over\alpha}.
\eeq
The key caveat is that the resulting plot shows not
the true $P(k)$ but a smoothed version thereof, 
with a roughly constant smoothing width 
$\Delta\ln k=\gamma$ on our logarithmic $k$-axis.

For the SDSS selection functions described in Dodelson {\etal} (2001), 
$\aa\approx 0.75$, $\bb\approx 1.1$ and the smoothing width 
$\cc=\Delta k/k\approx 0.37$ for all four magnitude bins.
Even major changes in the functional form of the radial selection
function do not change these shape parameters by large amounts.
In contrast, the mean survey depth varies substantially,
with 
$r_*=515h^{-1}$Mpc,
$710h^{-1}$Mpc,
$947h^{-1}$Mpc and 
$1193h^{-1}$Mpc for the four magnitude bins, respectively,
with non-negligible uncertainty (Dodelson {\etal} 2001).
  
Since the relative windows widths $\Delta k/k$ are so large, 
it is important to use \eq{kWinDefEq} rather than 
\eq{CPeq} on the largest scales, where $k^3 P(k)$ is far
from constant. Moreover, the
additional smearing $\Delta\l\sim 20$ caused by our finite sky
coverage becomes important on large scales, since it corresponds to
a rock-bottom smoothing scale $\Delta k\approx\bb\Delta\l/r_*$ that does
not decrease with $k$.

Perhaps the most useful feature of \eq{CPeq} is that it explicitly 
shows the effect of changing the radial selection function.
If the shape $f_*$ has been correctly estimated but the
mean survey depth $r_*$ has been overestimated, then
\eq{CPeq} shows that the inferred power spectrum $P(k)$ will 
be too far up to the left
--- up because there was in fact less averaging along the line of sight
suppressing the observed power $C_\l$, to the left because
a given angular scale $\l$ in fact corresponds to larger 
spatial scale. 
Any such errors will therefore slide the entire $P(k)$ curve
along the solid lines of slope $-3$ shown in \fig{Pfig}.

Although the dominant uncertainty is likely to arise from the mean depth
$r_*$, the dependence on the shape (as opposed to the mean depth)
of the selection function is also rather intuitive.
If the galaxies are more concentrated around their mean 
distance (if $f_*(x)$ is more sharply peaked around $x=1$),
then a straightforward calculation shows that
the normalization $\alpha$ increases and the 
smoothing width $\gamma=\Delta k/k$ decreases.
The first effect corresponds to less averaging down of fluctuations
along the line of sight, increasing $C_\l$ for a given $P(k)$.
The second effect corresponds to less aliasing, since 
the relation between angular separation and transverse spatial separation 
tightens when the galaxies become less spread out radially.
Since $\gamma$ is essentially the width of the 
the radial selection function in units of the mean depth
(more precisely, this ratio for the {\it square} of the selection function,
which is more peaked and therefore gives a smaller number),
it is difficult to obtain smearing $\gamma\Delta /k \simlt 25\%$
for realistic selection functions. However, much smaller
$\gamma$-values of course become possible if photometric redshifts
are used to define galaxy samples in narrow radial bins.


\end{document}